\documentclass{aa}  

\usepackage{graphicx,amsmath,wasysym}
\usepackage{txfonts}
\usepackage{amssymb}
\usepackage{color}
\usepackage{natbib}
\usepackage{url}
\usepackage{multirow}
\usepackage{threeparttable}
\bibpunct{(}{)}{;}{a}{}{,}
\usepackage{sidecap}
\usepackage{array}
\usepackage{tabularx}
\usepackage{txfonts}
\usepackage[version=3]{mhchem}
\defcitealias{Zhang2017}{ZS}

\def\approxinf{
  \def\p{
    \setbox0=\vbox{\hbox{$<$}}
    \ht0=0.6ex \box0 }
  \def\s{
    \vbox{\hbox{$\sim$}}
  }
  \mathrel{\raisebox{0.7ex}{
      \mbox{$\underset{\s}{\p}$}
    }}
}

\def\approxsup{%
  \def\p{%
    \setbox0=\vbox{\hbox{$>$}}%
    \ht0=0.6ex \box0 }%
  \def\s{%
    \vbox{\hbox{$\sim$}}%
  }%
  \mathrel{\raisebox{0.7ex}{%
      \mbox{$\underset{\s}{\p}$}%
    }}%
}

\begin{document}

   \title{Optical phase curve of the ultra-hot Jupiter WASP-121b}

   \author{
   V. Bourrier\inst{1},
   D. Kitzmann\inst{2},     
   T. Kuntzer\inst{1},
   V. Nascimbeni\inst{3},   
   M. Lendl\inst{1},         
   B. Lavie\inst{1},     
   H.J. Hoeijmakers\inst{1,2},    
   L. Pino\inst{4},    
   D. Ehrenreich\inst{1},   
   K. Heng\inst{2},         
   R. Allart\inst{1},  
   H.M Cegla\inst{1},    
   X. Dumusque\inst{1},   
   C. Melo\inst{5},    
   N. Astudillo-Defru\inst{6},
   D.A. Caldwell\inst{7,8},      
   M. Cretignier\inst{1},
   H. Giles\inst{1}, 
   C.E. Henze\inst{8},        
   J. Jenkins\inst{8}, 
   C. Lovis\inst{1}, 
   F. Murgas\inst{9,10}, 
   F. Pepe\inst{1}, 
   G.R. Ricker\inst{11},        
   M.E. Rose\inst{8},          
   S. Seager\inst{11,12,13}, 
   D. Segransan\inst{1}, 
   A. Su\'arez-Mascare\~no\inst{9},
   S. Udry\inst{1},  
   R. Vanderspek\inst{11},
   A. Wyttenbach\inst{14},    
          }

\authorrunning{V.~Bourrier et al.}
\titlerunning{Phase curve of WASP-121b}

\offprints{V.B. (\email{vincent.bourrier@unige.ch})}

\institute{
Observatoire de l'Universit\'e de Gen\`eve, 51 chemin des Maillettes, 1290 Versoix, Switzerland     
\and 		
Center for Space and Habitability, Universit\"at Bern, Gesellschaftsstrasse 6, 3012 Bern, Switzerland     
\and
INAF – Osservatorio Astronomico di Padova, Vicolo dell’Osservatorio 5, 35122, Padova, Italy     
\and
Anton Pannekoek Institute for Astronomy, University of Amsterdam, Science Park 904, 1098 XH Amsterdam, The Netherlands       
\and
European Southern Observatory, Alonso de C\'ordova 3107, Vitacura, Región Metropolitana, Chile      
\and        
Departamento de Matem\'atica y F\'isica Aplicadas, Universidad Cat\'olica de la Sant\'isima Concepci\'on, Alonso de Rivera 2850, Concepci\'on, Chile         
\and
SETI Institute, Mountain View, CA 94043, USA     
\and
NASA Ames Research Center, Moffett Field, CA 94035, USA      
\and
Instituto de Astrofísica de Canarias (IAC), E-38205 La Laguna, Tenerife, Spain    
\and       
Departamento de Astrofísica, Universidad de La Laguna (ULL), E-38206 La Laguna,   Tenerife, Spain   
\and
Department of Physics and Kavli Institute for Astrophysics and Space Research, MIT, Cambridge, MA 02139, USA      
\and
Department of Earth, Atmospheric and Planetary Sciences, Massachusetts Institute of Technology, Cambridge, MA 02139, USA   
\and
Department of Aeronautics and Astronautics, MIT, 77 Massachusetts Avenue, Cambridge, MA 02139, USA   
\and
Leiden Observatory, Leiden University, Postbus 9513, 2300 RA Leiden, The Netherlands   
}

   \date{}

 \abstract{We present the analysis of TESS optical photometry of WASP-121b, which reveal the phase curve of this transiting ultra-hot Jupiter. Its hotspot is located at the substellar point, showing inefficient heat transport from the dayside (2870\,K) to the nightside ($<$ 2200\,K) at the altitudes probed by TESS. The TESS eclipse depth, measured at the shortest wavelength to date for WASP-121b, confirms the strong deviation from blackbody planetary emission. Our atmospheric retrieval on the complete emission spectrum supports the presence of a temperature inversion, which can be explained by the presence of VO and possibly TiO and FeH. The strong planetary emission at short wavelengths could arise from an H$^{-}$ continuum.}


   \maketitle
%

\section{Introduction}

The discovery of hot Jupiters opened a window into planetary atmospheres shaped by extreme irradiation conditions not found in the solar system. Some of these giant planets are on such close orbits around their star that their dayside temperature is raised to more than 2000\,K (\citealt{Parmentier2018}), facilitating the measurement of their thermal emission (e.g. \citealt{Arcangeli2018}) and simplifying their atmospheric chemistry (\citealt{Lothringer2018}). 

The ultra-hot Jupiter WASP-121b (\citealt{Delrez2016}) is a good candidate for atmospheric studies. This super-inflated gas giant transits a bright F6-type star (V = 10.4, J = 9.6), favoring optical and infrared emission spectroscopy measurements. Infrared spectroscopy with the Hubble Space Telescope revealed the presence of a thermal inversion, via the resolved emission signature of water in the planet dayside (\citealt{Evans2017}). High-altitude absorbers like vanadium and titanium oxydes have been proposed to explain the formation of this stratosphere (\citealt{Evans2017}). Additional single-band measurements of  WASP-121b secondary eclipse (\citealt{Delrez2016}, \citealt{Kovacs2019}) hint at the departure of the planetary dayside emission from an isothermal blackbody, and bring further constraints on the atmospheric composition. Transmission spectroscopy at optical and infrared wavelengths further showed the signature of water in absorption at the atmospheric limb, and the possible presence of vanadium oxide and iron hydride, with no titanium oxide (\citealt{Evans2016,Evans2018}). Alternative species could, however, explain features in the emission and transmission spectra (e.g., \citealt{Parmentier2018}, \citealt{Gandhi2019}). 

In the present study, we aim at extending our understanding of the thermal emission and atmospheric structure of WASP-121b. We present the analysis of TESS photometry of WASP-121 in Sect.~\ref{sec:ph_curve}, along with the interpretation of the planetary phase curve, primary transit and secondary eclipse. In Sect.~\ref{sec:atm_dayside}, we characterize the atmospheric structure of the planet on the dayside, via the analysis of its emission spectrum. We conclude the study in Sect.~\ref{sec:conclu}.

\section{TESS photometry of the WASP-121 system}
\label{sec:ph_curve}

\subsection{Preprocessing}

\begin{figure*}
\begin{center}
\includegraphics[width=1.\linewidth]{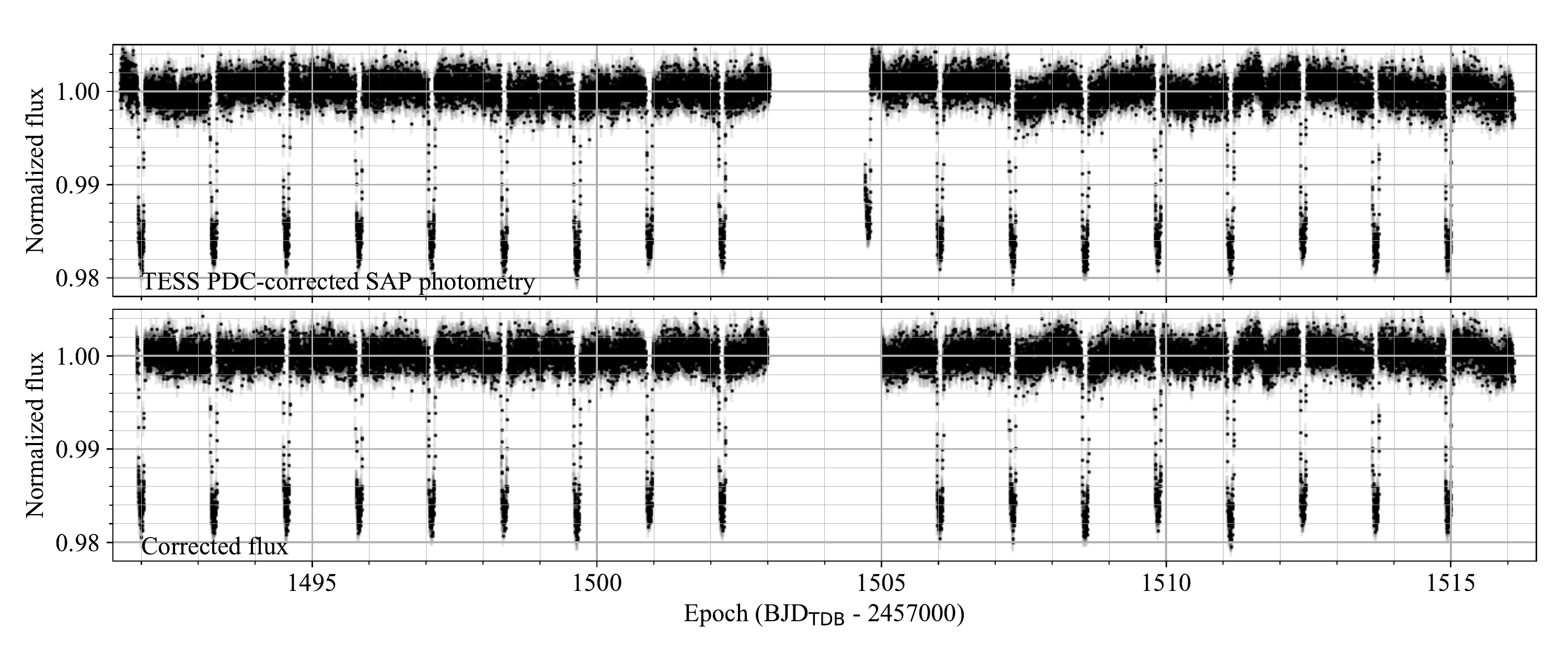}
\caption{\label{fig:TESS:lc} (\emph{Top panel}) Normalized TESS PDC-corrected photometry of WASP-121, with error bars, obtained over TESS orbits 21 and 22. (\emph{Bottom panel}) PDC flux after correcting by the median per orbital period of WASP-121b, and removing ramp-like systematics at the start of each TESS orbit (see text for details). We used this time series for the analysis.}
\end{center}
\end{figure*}

WASP-121 (also known as TIC 22529346) was observed by the TESS \citep{Ricker2015} mission in sector 7, camera 3. Short cadence (two-minute) data were acquired over two TESS orbits (21 and 22) between 08 January and 01 February 2019, spanning 24.5 days and covering 18 primary transits of WASP-121b. We retrieved the photometry generated by the TESS Science Processing Operation Center (SPOC), which provides the simple aperture photometry (SAP) and a Presearch Data Conditioning flux (PDC) \citep{Jenkins2016}. The latter algorithm works in a similar way as the Kepler Presearch Data Conditioning algorithm \citep{Stumpe2012, Smith2012}, which corrects the SAP photometry for instrumental effects. The median-normalized PDC photometry is presented in the upper panel of Fig~\ref{fig:TESS:lc}. The baseline flux shows a ramp-like decrease at the start of each TESS orbit, as expected from previous TESS observations (eg \citealt{Shporer2019}). We thus excluded the measurements obtained before 1491.92 (BJD - 2457000) in orbit 21 and before 1505.00 in orbit 22. We then apply a median-detrending algorithm, with a window size of one orbital period of WASP-121b, to remove remaining systematics while keeping variability at the planetary period intact. The corrected photometry used in our analysis is shown in the lower panel of Fig.~\ref{fig:TESS:lc}. 

The light curve of WASP-121 generated a TESS alert identified as TOI-495. The automated alert pipeline detected WASP-121b (TOI-495.01) but also a candidate second planet (TOI-495.02) via two transits events at a period of 19.09 days. We first excluded these events from our analysis of WASP-121b. Phase-folding the photometry corrected for the model signal of WASP-121b then showed unphysical drops in flux at the time of the candidate transits. Furthermore, there are no indications for a planet with this period in \citet{Delrez2016}. Additional photometry and RV data with adequate sampling are required to assess the nature of TOI-495.02.\\

\subsection{Model}
\label{sec:TESS:mcmc}

\begin{figure*}
\begin{center}
\includegraphics[width=1.\linewidth]{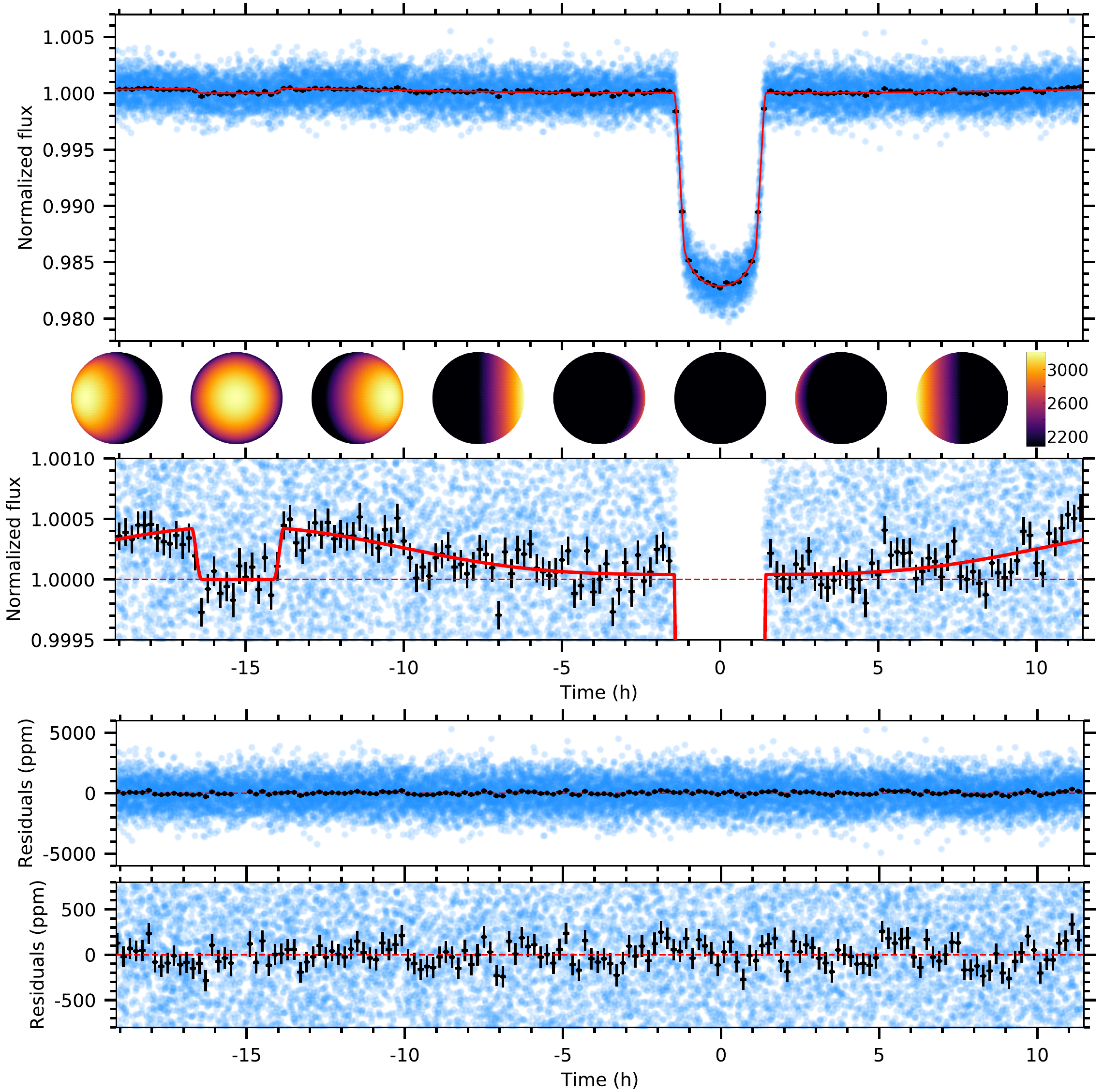}
\caption{\label{fig:TESS:fit} TESS light curve of WASP-121b. \textit{Upper panels}: Photometric data (blue points) phase-folded at the orbital period of the planet. The lower sub-panel zooms in on the planetary phase curve, clearly visible in the binned exposures (black points). Our best-fit model to the complete light curve is plotted as a solid red line. The corresponding temperature distribution of the planet is shown at regular phase intervals. The flux is normalized to unity during the secondary eclipse, when the stellar light alone is measured.  \textit{Lower panels}: Residuals between the photometric data and the best-fit.}
\end{center}
\end{figure*}

Transits of WASP-121b were previously measured from the near-ultraviolet to the near-infrared by \citet{Delrez2016}, \citet{Evans2016,Evans2017,Evans2018}, and \citet{Salz2019_NUV_WASP121b}. Secondary eclipses were further unveiled in WASP-121 optical and near-infrared photometry by \citet{Delrez2016}, \citet{Evans2017,Evans2019}, \citet{Kovacs2019}, and \citealt{Garhart2019}. By-eye inspection of our corrected TESS photometry, folded at the orbital period of WASP-121\,b and binned (Fig.~\ref{fig:TESS:fit}), clearly shows the primary transit and secondary eclipse. It further reveals the phase curve of this planet.

We fitted a global model to the TESS light curve to recover the properties of WASP-121b and its star. The Python packages \texttt{batman} \citep{kreidberg2015} and \texttt{spiderman} \citep{Louden2018} were combined to respectively model the transit, and the phase curve modulation with the secondary eclipse. The light curve was modeled as:
\begin{equation}
F(t) = C_{\rm s} \cdot F_{*} \cdot \left(  \delta_{\rm tr}(t) + \frac{F_{\rm p}^{\rm thermal}(t)}
{F_{*}} \right)  
\label{eq:eq_LC}
\end{equation}
The scaling coefficient $C_{\rm s}$ is used to set the stellar flux to unity during secondary eclipse. The relative flux variation $\delta_{\rm tr}(t)$ due to the transit of WASP121b is calculated with \texttt{batman}, assuming quadratic limb darkening. The baseline stellar flux $F_{*}$ is calculated with a Phoenix spectrum representative of WASP-121 (T$_{*}$ = 6400\,K, log(g) = 4.24 , [Fe/H] = 0.13; \citealt{Husser2013}), and integrated over the TESS band (600 to 1000\,nm, \citealt{Ricker2015}). The planetary flux is defined relative to the stellar flux in Eq.~\ref{eq:eq_LC} because this is how \texttt{spiderman} calculates the phase curve with the secondary eclipse. We used a semi-physical brightness map based on \citet[][ZS in the following]{Zhang2017} to approximate the planet, assuming that its emission $F_{\rm p}^{\rm thermal}(t)$ is purely thermal. This model requires three planet-dependent parameters: 
\begin{enumerate}
 \item The ratio of radiative to advective timescales, $\xi$. The advective timescale, $\tau_\text{adv}$, is the ratio of a typical length of the system divided by the zonal-mean zonal wind. The $\xi$ parameter controls the longitudinal position of the maximum temperature on the planet. If $\xi$ is close to 0, the maximum temperature is close to the sub-stellar point and is measured near the phase of the secondary eclipse. We note that the advective timescale can take negative values, corresponding to winds going from east to west and shifting the hot spot westward of the sub-stellar point ($\xi \approxinf$ 0).
 \item The radiative equilibrium temperature of the nightside, $T_N$. If $\xi$ is close to 0, the temperature distribution of the nightside is uniform and equal to $T_N$. As $\xi$ increases, advection redistributes the heat more efficiently across longitudes until the temperature distribution becomes uniform. For large $\xi$, the average temperature of the nightside (which controls the planetary flux measured in excess of the stellar flux, when the planet nightside is fully visible) can thus be higher than $T_N$.
 \item The contrast of day-night radiative equilibrium temperatures, $\Delta T_{DN}$, defined between the anti-stellar and sub-stellar points. We caution that for large $\xi$, the temperature of these points will be respectively larger and lower (corresponding to a lower contrast) than the purely radiative temperatures. The amplitude of the phase curve is mostly controlled by $T_N$ and $\Delta T_{DN}$. 
\end{enumerate}

We fitted the global model to the TESS photometry time series by running the \texttt{emcee} \citep{Foreman2013} Markov Chain Monte Carlo (MCMC) algorithm. The model is oversampled and averaged within the 2-min windows of the TESS exposures. Model parameters were used as jump parameters, replacing the orbital inclination by its cosine, and the limb-darkening coefficients by the linear combinations $c_{1}$ = 2\,$u_{1}$ + $u_{2}$ and $c_{2}$ = $u_{1}$ - 2\,$u_{2}$ (\citealt{Holman2006}). The chosen priors are given in Table~\ref{tab:TESS:priors}. We ran 200 walkers over 3000 steps, and removed 500 burn-in steps. We checked that all walkers converged toward the same solution before merging the chains and calculating the posterior distributions for the model parameters.

\subsection{Results}
\label{sec:TESS:analysis}

The results of our fit are given in Table~\ref{tab:sys_prop}, with the corresponding model displayed in Fig.~\ref{fig:TESS:fit}. It provides a good fit to the TESS photometry, yielding a reduced $\chi^2$ = 0.98 (13 free parameters, 15773 degrees of freedom). Correlation diagrams for the probability distributions of the model parameters are shown in Fig.~\ref{fig:WASP121_triangle}. Best-fit values for the model parameters were set to the median of their distributions, except for the eccentricity whose distribution favors a circular orbit. Its best-fit value was set to 0, and we provide upper limits at 1 and 3$\sigma$ in Table~\ref{tab:sys_prop}. Some of the other parameter distributions are asymmetrical, and we thus chose to define their $1\sigma$ uncertainties using the highest density intervals, which contains 68.3\% of the posterior distribution mass such that no point outside the interval has a higher density than any point within it. 

Overall our results for the orbital and transit properties are consistent with those of the TESS alert pipeline and previous analyses of the system. \citet{Evans2018} noted a discrepancy between their HST-derived values for a$_\mathrm{p}/$R$_{*}$ and $b$ and the ones previously derived from ground-based measurements by \citet{Delrez2016}. Our values are in between those from these two studies, and only discrepant with the a$_\mathrm{p}/$R$_{*}$ from \citet{Delrez2016} at 2.8$\sigma$. We substantially refine the precision on the orbital eccentricity (\citealt{Kovacs2019}) and confirm that WASP-121b is on a circular orbit with e $<$ 0.0078 (3$\sigma$). WASP-121b is close enough to its star that extreme tidal forces are expected to induce significant departure from sphericity in the planet. \citet{Akinsanmi2019} estimated that a precision of 50 ppm/min noise level will be necessary to measure this deformation, and gain information on the interior structure of the planet. Unfortunately, this precision remains out of reach of the current TESS data (Fig.~\ref{fig:TESS:fit}), justifying our use of a spherical transit model. 

Significant temporal variations were recently measured in the optical occultation depth of the ultra-hot Jupiter WASP-12b (\citealt{VonEssen2019}, \citealt{Hooton2019}), possibly tracing variable thermal emission in the dayside atmosphere, changes in the cloud coverage, or scattering by escaping planetary material. Modelling independently the TESS photometry of WASP-121b in each orbital revolution, we found the individual transit and occultation depths to be well spread around the global posterior estimates, except for a few values coinciding with epochs of lesser pointing accuracy of the TESS telescope (\citealt{Fausnaugh2019}). We thus conclude to the non-detection of temporal variations in the transit and occultation depths of WASP-121b.  \\

\subsection{Temperature distribution}
\label{sec:TESS_Temp}

A fraction of the light coming from WASP-121b could come from reflected starlight, besides the planetary thermal emission. To check the validity of our assumptions, we added a reflection component to the \citetalias{Zhang2017} thermal model. The planetary atmosphere is approximated by \texttt{spiderman} as a Lambertian sphere characterized by a geometric albedo $A_{\rm g}$ and reflecting evenly at all phase angles. When fitting the data without any a-priori assumptions about the planetary temperature, we found that the MCMC favors a purely reflecting solution, with no thermal component and $A_{\rm g}$ = 0.37$\stackrel{+0.03}{_{-0.04}}$. A similar value $A_{\rm g}$ = $\delta (a_{\rm p}/R_{\rm p})^{2}$ = 0.39$\pm$0.04 is derived from the TESS eclipse depth ($\delta$ = 419$\stackrel{+47}{_{-42}}$\,ppm). This scenario however yields a similar $\chi^2$ as the purely thermal solution derived previously and a larger Bayesian information criterion (BIC, \citealt{Schwarz1978}, \citealt{Liddle2007}), showing that the TESS data do not contain enough information to include a reflection component. Assuming that scattering by atmospheric aerosols is isotropic, the corresponding Bond albedo of 0.56 would also be atypical for a planet like WASP-121b (it falls in the category of class V giants at small orbital distance around an F-type star in \citealt{Sudarsky2000}, with maximum theoretical Bond albedos of 0.56). Optical geometric albedos derived from Kepler data of 11 hot Jupiters were found to typically range between 0 and 0.2, with only one higher value of 0.352 for Kepler-7b (\citealt{Heng2013}; see also the case of $\tau$ Bootis b, with $A_{\rm g} < $ 0.12 at 3$\sigma$, \citealt{Hoeijmakers2018a}). Analysis of eclipse depths around 0.9$\mu$m by \citet{Mallonn2019} consistently point to a low reflectivity in the optical to near-infrared transition regime for hot to ultra-hot Jupiters (with a 3$\sigma$ upper limit of 0.37 on $A_{\rm g}$ for WASP-121b). A geometric albedo of 0.37 for WASP-121b thus appears unlikely, especially considering the difficulty to form condensates at $\sim$3000\,K (see Sect.~\ref{sec:atm_dayside}, \citealt{Parmentier2018}; \citealt{Wakeford2017}). We thus favour hereafter the scenario where the optical planetary flux of WASP-121b is dominated by thermal emission.\\

The ratio of radiative to advective timescales derived from the fit to the TESS phase curve, $\xi$, is consistent with zero. This implies a poor heat redistribution between the day and night sides of WASP-121b, with the hotspot located at the sub-stellar point and the atmospheric thermal structure dominated by radiation in the layers probed by TESS. This result is in line with theoretical expectations that the hotspot offset should decrease as the irradiation of the planet increases (\citealt{Perna2012}, \citealt{Parmentier2018b}). On the planet dayside the best-fit model yields a maximum temperature $T_{\rm N}$ + $\Delta T_{\rm DN}$ = 3225$^{+65}_{-88}$\,K at the sub-stellar point, well-constrained by the eclipse depth. With $\xi$ close to 0, the model dayside temperature varies with longitude $\theta_{\rm long}$ as $T_{\rm N}$ + $\Delta T_{\rm DN}$\,cos($\theta_{\rm long}$) (\citetalias{Zhang2017}). Integrating over the dayside yields an average temperature of 2794$^{+101}_{-95}$\,K. On the nightside, the low $\xi$ value results in a uniform temperature distribution equal to the derived radiative temperature ($T_\mathrm{N}$ = 2082$^{+338}_{-271}$\,K). However, we note that the planet-to-star flux ratio measured directly before ingress/after egress is only marginally detected (63$\pm$27\,ppm), preventing a clear detection of WASP-121b nightside emission. In the \citetalias{Zhang2017} model the temperature depends on $T_\mathrm{N}$ at all longitudes, and its value is thus constrained here by the shape of the phase curve rather than by the nightside flux. 

We compared these model-dependent values to the measurements of the planetary flux derived directly from the phase curve. We used the Phoenix spectrum representative of WASP-121 (Sect.~\ref{sec:TESS:mcmc}) to extract the planetary flux from the eclipse depth, which was then fitted with a blackbody model integrated over the TESS passband. This yields a planetary brightness temperature $T_\mathrm{D}^\mathrm{T}$ = 2870$\pm$50\,K, which corresponds to the average dayside emission in the TESS band and is in good agreement with the value from the \citetalias{Zhang2017} best-fit model. The blackbody fit to the excess planetary flux before ingress/after egress yields a temperature $T_\mathrm{N}^\mathrm{T}$ = 2190$^{+106}_{-145}$\,K, consistent with the model result but also with zero at 2$\sigma$ (ie, a non-detection of the nightside emission). This comparison nonetheless supports the validity of the \citetalias{Zhang2017} kinematic model for WASP-121b, and the dominance of radiation over advection for this planet, in agreement with previous analyses at longer wavelengths  \citep{Delrez2016, Evans2017, Kovacs2019}. The brightness temperature we derive for the dayside is consistent with those derived by \citealt{Evans2019} from HST eclipse depths between 0.8 and 1$\mu$m. This suggests that significant molecular dissociation occurs on the dayside of WASP-121b, which should have comparable amounts of H$^{-}$ and H$_{2}$, or even be H$^{-}$-dominated (see \citealt{Parmentier2018} and \citealt{Evans2019} for a similar conclusion from the analysis of the dayside emission spectrum). In contrast the lower temperature of the nightside, resulting from poor heat redistribution and low wind speeds from the dayside, implies that it is likely dominated by H$_{2}$ at the altitudes probed by TESS (\citealt{Bell2018}, \citealt{Kitzmann2018}).

\begin{table*}   
\caption[]{Properties for the WASP-121 system}
\centering
\begin{threeparttable}
\begin{tabular}{c|c|c|c}
    \hline
    \hline
      Parameter & Symbol & Value  &  Unit    \\
    \hline      
\textit{Stellar properties}    &   &   &    \\      
    \hline   
Mass		& $M_{\star}$ & 1.358$\stackrel{+0.075}{_{-0.084}}$   & M$_{\odot}$   \\
Radius		& $R_{\star}$ & 1.458$\pm$0.030  & R$_{\odot}$   \\    
Density$^{\dagger}$		& $\rho_{\star}$ & 0.434$\pm$0.038  &  $\rho_{\odot}$  \\
Limb-darkening coefficients & $u_{1}$ & 0.268$\pm$0.039   &  \\
			& $u_{2}$ & 0.138$^{+0.075}_{-0.078} $ &     \\
    \hline
    \hline 
\textit{Planetary properties}    &   &    \\ 
    \hline
Transit epoch	& $T_{0}$ & 2458119.72074$\pm$0.00017 &   BJD$_\mathrm{TDB}$  \\    
Orbital period & $P$ & 1.27492485$\pm$5.6$\times$10$^{-7}$  &  d  \\  
Scaled semi-major axis & $a_\mathrm{p}/R_{\star}$ & 3.8216$^{+0.0078}_{-0.0084}$ &   \\
Semi-major axis$^{\dagger}$ & $a_\mathrm{p}$ & 0.02591$\pm$0.00054 &  au   \\			
Eccentricity & $e$ & [ 0 - 0.0032 ]  [ 0 - 0.0078 ]  &   \\
Argument of periastron & $\omega$ & 10$\pm$10 &   deg    \\
Orbital inclination 	& $i_\mathrm{p}$ & 89.10$\stackrel{+0.58}{_{-0.62}}$ &  deg   \\     
Impact parameter$^{\dagger}$ & $b$ & 0.060$\stackrel{+0.041}{_{-0.039}}$ &       \\ 
Transit durations$^{\dagger}$ & $T_\mathrm{14}$ &  2.9059$\stackrel{+0.0062}{_{-0.0057}}$   &   h   \\
				& $T_\mathrm{23}$ &   2.2639$\pm$0.0051    &   h  \\
Planet-to-star radii ratio & $R_\mathrm{p}/R_{\star}$ & 0.12355$\stackrel{+0.00033}{_{-0.00029}}$ &    
\\  
Radius$^{\dagger}$      & $R_\mathrm{p}$ &  1.753$\pm$0.036 & $R_\mathrm{Jup}$   \\
Planet-to-star flux ratios & $F^\mathrm{T}_\mathrm{p}(\rm day)/F_{\star}$ & 419$^{+47}_{-41}$ & ppm    
\\  
      & $F^\mathrm{T}_\mathrm{p}(\rm night)/F_{\star}$ & 63$\pm$27 & ppm    
\\  
Nightside temperature  & $T_\mathrm{N}^\mathrm{T}$ &  2190$^{+106}_{-145}$  &   K    \\
Dayside temperature  & $T_\mathrm{D}^\mathrm{T}$ &  2870$\pm$50  &   K    \\
Radiative to advective timescales ratio & $\xi$ &  -0.016$\stackrel{+0.061}{_{-0.064}}$ &      \\	 			       
    \hline
  \end{tabular}
  \begin{tablenotes}[para,flushleft]
  Notes: The mass, radius, and density of the star come from \citealt{Delrez2016}. All other properties come from the present work. Values in brackets for the eccentricity indicate the 1 and 3$\sigma$ confidence intervals with the lower limit set a 0. Coefficients $u_1$ and $u_2$ are associated with a quadratic limb-darkening law. $^{\dagger}$ indicate derived parameters. The planet brightness temperatures are derived directly from the planet-to-star flux ratios measured in the TESS band. \\
  \end{tablenotes}
  \end{threeparttable}
\label{tab:sys_prop}
\end{table*}



\section{Atmospheric dayside structure}
\label{sec:atm_dayside}

\begin{figure}
\centering
\includegraphics[trim=0cm 0.cm 0.cm 0.cm,clip=true,width=\columnwidth]{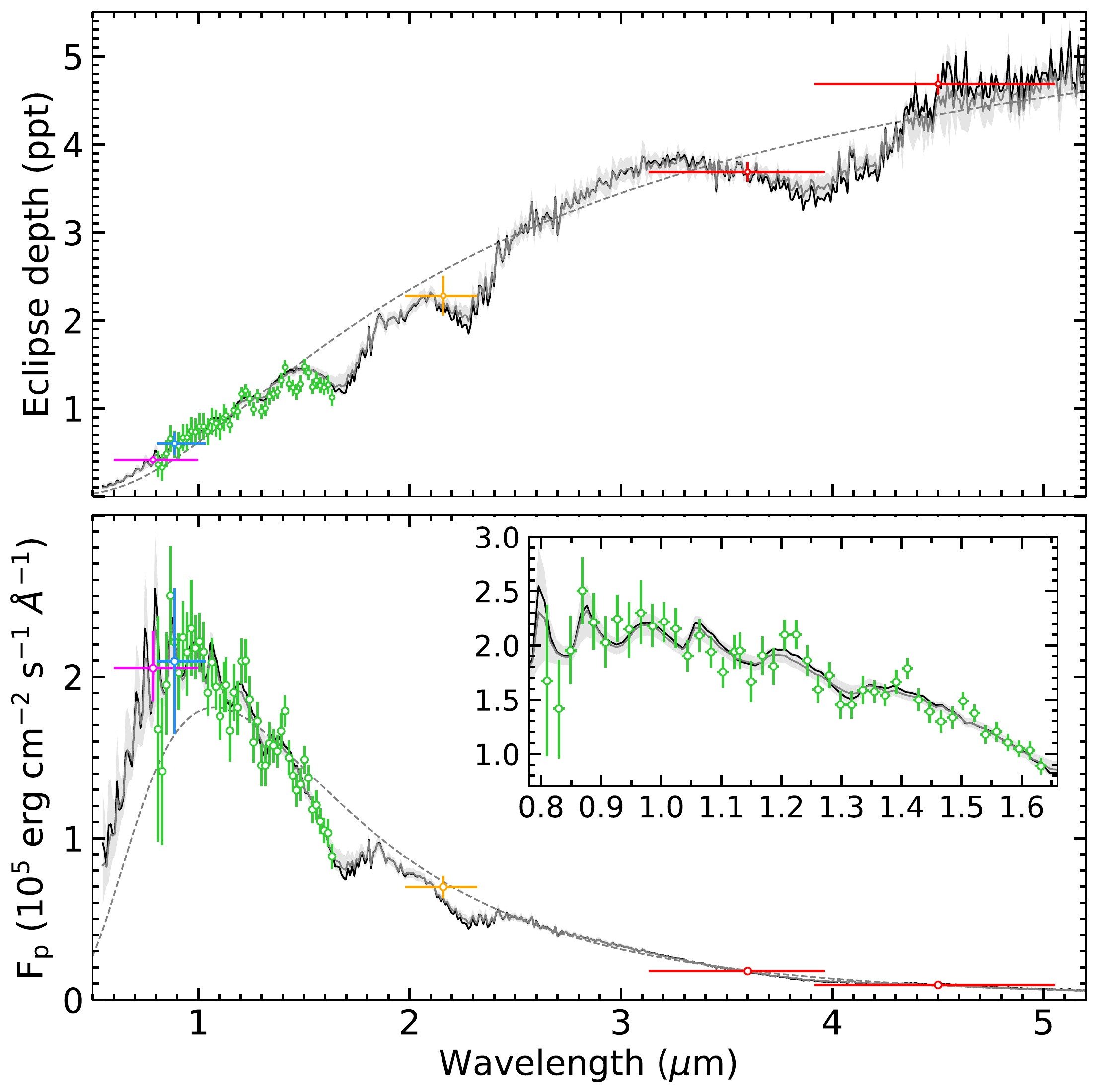}
\caption[]{Eclipse depths and emission spectrum of WASP-121b at optical and infrared wavelengths. \textit{Top panel}: Planet-to-star flux ratios measured with TESS (purple), TRAPPIST (Sloan-z' band, blue), HST (WFC3, green), SMARTS’ (2MASS K band,  orange), and Spitzer (IRAC, red). Horizontal bars indicate the bandpasses of the instruments. \textit{Bottom panel}: Corresponding planetary emission spectrum, with flux defined at the TESS planet radius. In both panels, the grey dashed line shows the best-fit isothermal blackbody spectrum, and the black solid line the best-fit model from our retrieval analysis. The shaded region is the 1-$\sigma$ envelope of all spectra derived from the posterior distributions, and the grey solid line is the corresponding median spectrum. For the sake of clarity all original high-resolution spectra have been binned down to a lower resolution. The inset plot shows a magnification of the WFC3 wavelength range. }
\label{fig:em_atm}
\end{figure}

The TESS eclipse depth is the bluest value obtained for WASP-121b, completing measurements obtained with TRAPPIST/Sloan-z' (\citet{Delrez2016}), SMARTS’/2MASS K (\citealt{Kovacs2019}), HST/WFC3 (\citet{Evans2017,Evans2019}), and Spitzer/IRAC (\citet{Evans2017}, \citealt{Garhart2019}). The full eclipse depths and planetary emission spectra are displayed in Fig.~\ref{fig:em_atm}. Eclipse depths were converted into planetary fluxes using the same approach as in Sect.~\ref{sec:TESS_Temp}. Fitting a black body to the planetary emission spectrum yields a poor fit (reduced $\chi^{2}$ = 3.4) with a spectrally-averaged dayside temperature of 2690$\pm$7\,K. An isothermal blackbody spectrum does not fit well the features in the HST/WFC3 data (see also \citealt{Evans2017, Evans2019}), and cannot explain both the optical and infrared measurements. This is consistent with the lack of efficient heat transport from the dayside to the nightside revealed by the TESS planetary phase curve. We thus performed a full retrieval analysis of WASP-121b emission spectrum to better interpret the data and gain insights into the planet dayside chemical composition and thermal structure.\\

\subsection{Retrieval from the emission spectrum}

For this analysis, we employ the new version of the \texttt{Helios-r} retrieval code \citep{Lavie2017_helios}. \texttt{Helios-r.2} still uses a nested sampling algorithm (\citealt{skilling_2006}, \citealt{feroz_multimodal_2008,feroz_multinest_2009}) to sweep the parameter space of the atmospheric model, which includes improvements over the previous version (Kitzamann et al. 2019 in prep). The temperature profile can be retrieved freely within \texttt{Helios-r.2} and is not tied to any pre-determined functional form. We base our description of the temperature profile on the theory of finite element methods. More specifically, we partition the atmosphere into a given number of non-overlapping elements, distributed equidistantly in the log pressure space. Within each of these elements, the temperature profile is approximated by a piecewise polynomial of a given degree, while additionally forcing the temperatures to be continuous across element boundaries. Further descriptions and details of \texttt{Helios-r.2} are given in Kitzmann et al. (2019, in prep.). 

For the retrieval of WASP-121b, we perform a free chemistry retrieval. The retrieved molecules' abundances are assumed to be constant throughout the atmosphere and can be interpreted as the mean value of the species abundance near the photosphere of the planet. Based on the study of \citet{Evans2018}, we include \ce{H2O}, \ce{CO}, \ce{CH4}, \ce{VO}, \ce{FeH}, and \ce{TiO} in our retrieval analysis. Additionally, we also add \ce{H-} because of its strong continuum contribution in the HST/WFC3, TESS, and TRAPPIST bandpasses. The presence of \ce{H-} has been previously proposed by \citet{Parmentier2018}. It is expected from the high temperatures in the atmosphere of WASP-121b \citep{Kitzmann2018}, and supported by our fit to the planet optical phase curve (Sect.~\ref{sec:TESS_Temp}). We also performed tests with additional chemical species, such as \ce{NH3}, or the alkali metals K and Na, all of which were not favoured by the Bayesian framework. We expect that Al, Ca, and Mg (\citealt{Gandhi2019}) will mostly be locked up in condensates in the deeper atmosphere. We also do not expect Fe to be present in atomic form at the high pressure / low temperature levels probed by the emission spectrum (\citealt{Lothringer2018}). Consequently, we neglect all these species in the final retrieval. We assume that the rest of the atmosphere is made up of \ce{H2} and He, with their mixing ratios being determined by the ratio of their solar elemental abundance.

For the temperature profile, we use four second-order elements from 100 bar to $10^{-6}$ bar. The only directly retrieved temperature is the one at the bottom of the atmosphere, for which we use a uniform prior between 5500 and 2000\,K. For all subsequent temperature points, we choose to use the temperature's lapse rate as the basis of our temperature retrieval. Thus, we introduce parameters $b_i$, such that the temperature at a point $i$ is given by:
\begin{equation}
 T_i = b_i T_{i-1} .
\end{equation}
Since we expect the temperature profile to have an inversion somewhere in the atmosphere (\citealt{Evans2017}), we use normal priors between 0.5 and 1.5 for the $b_i$ parameters. For the four second-order elements, we thus have one free temperature at the bottom and eight values $b_i$, for a total of nine free parameters to describe the temperature profile.

Additionally, we also add the measured surface gravity and the ratio $\left(R_\mathrm{s}/R_\mathrm{p}\right)^2$ to the retrieval, where $R_\mathrm{s}$ and $R_\mathrm{p}$ are the stellar and planetary radius, respectively. This value is required to convert the eclipse depths into the planet's actual emission spectrum. For these parameters, we use Gaussian priors centred on their respective measured value, with a standard deviation equal to the errors stated in Table~\ref{tab:sys_prop}. These priors propagate the errors of the inferred $\log g$ values and measured radii to the other retrieval parameters. Together with the abundances of the aforementioned molecules, we have in total 18 free parameters.

\subsection{Results}

The resulting posterior distributions and the obtained temperature profile are shown in Fig.~\ref{fig:retrival_posterior}. For each sample in the posterior distributions, a spectrum is calculated and then integrated within the bandpasses of the instruments used for the measurement of WASP-121b. The resulting median values are shown in Fig.~\ref{fig:em_atm}, together with the spectrum of the best-fit model, i.e. the parameter combination with the highest likelihood value. The results indicate that our retrieval is able to reproduce most of the measured data points. Except for a few data points, the median values or their confidence intervals are located within the error bars of the measurements. There are very few outliers within the WFC3 wavelength range that could be the result of an erroneous measurement \citep{Evans2019} or might indicate that our retrieval model does not include all details required to simulate the atmosphere of WASP-121b (e.g. a missing absorbing species). It should be noted that despite the rather wide 1-$\sigma$ limits of the retrieved atmospheric temperature profile (Fig.~\ref{fig:retrival_posterior}), the theoretical emission spectra show a quite small confidence interval (Fig.~\ref{fig:em_atm}). \\

Our results confirm those of \citet{Evans2017} and \citet{Evans2019} with respect to the presence of an inverted temperature profile in the atmosphere of WASP-121b. We also constrain the abundances of water, carbon monoxide, \ce{H-}, and VO. Within their confidence intervals, the derived \ce{H2O}, \ce{CO}, and \ce{H-} abundances are consistent with solar or slightly super-solar metallicities \citep[cf.][]{Kitzmann2018}. The VO mixing ratio, on the other hand, requires a super-solar abundance as also reported by \citet{Evans2019}. Our \ce{H2O} and \ce{VO} abundances are orders of magnitudes smaller than the ones reported by \citet{Evans2018}. It should, however, be noted that the VO abundance of $10^{-3.5}$ derived in \citet{Evans2018} is probably too high when considering realistic elemental abundances for vanadium \citep{Asplund2009}. In a follow-up study, \citet{Evans2019} remarked that the corresponding data points in the WFC3 spectrum that drive the VO detections are probably a statistical fluctuation. Since both VO, and \ce{H2O}, dominate the WFC3 bandpass, both quantities show a strong correlation in Fig.~\ref{fig:retrival_posterior}. The same also applies to the \ce{H-} abundance, as this anion has a strong continuum contribution in that wavelength range. In case of \ce{CH4}, FeH, and TiO, we obtain only upper limits for their abundances of about 1e-6. Methane is not expected to be present at large amounts due to the high temperatures \citep{Kitzmann2018}. The other two molecules are most likely strongly depleted by condensation below the planet's photosphere, which is consistent with the results reported in \citet{Evans2019}.\\

The median temperature profile and its confidence interval show a clear sign of an atmospheric temperature inversion (Fig.~\ref{fig:retrival_posterior}), probably caused by absorption of stellar radiation by shortwave absorbers, such as metal hydrides but also VO or TiO. In addition to the directly retrieved parameters, we also derive the corresponding planet's effective temperature. The posterior values for T$_\mathrm{eff}$ are obtained by calculating a high-resolution spectrum for each sample in the retrieved posterior distributions. Each spectrum is then integrated between 0.5 $\mu$m and 20 $\mu$m, and the resulting total flux converted into an effective temperature by using the Stefan-Boltzmann law. The effective temperature is strongly constrained, with a median of $2712 \pm 15$ K.

\section{Conclusions}
\label{sec:conclu}

Ultra-hot Jupiters magnify the gravitational and energetic interactions to which close-in planets are subjected to from their star. Their strongly irradiated atmospheres constitute excellent laboratories to study the dynamics and chemistry of close-in exoplanets. The ultra-hot Jupiter WASP-121b, transiting a bright F-type star on a near-polar orbit, is one such laboratory. 

TESS photometry reveals the secondary eclipse of WASP-121b at optical wavelengths and the planetary phase curve. It is consistent with pure thermal emission from a radiative atmosphere with inefficient heat redistribution, leading to a strong contrast ($\approxsup$700\,K) between the dayside and nightside brightness temperatures. The TESS eclipse depth extends the measured emission spectrum of WASP-121b from 4.5 to 0.8$\mu$m. Its interpretation with the Helios-r.2 retrieval code confirms the presence of a temperature inversion (\citealt{Evans2019}) and constrain the abundances of \ce{H2O}, \ce{CO}, \ce{H-}, and VO in the planet dayside atmosphere.

TESS, CHEOPS, and the JWST will enable the measurement of ultra-hot Jupiters phase curve from the optical to the mid-infrared domain, improving our understanding of their temperature distribution and global atmospheric circulation patterns. \\

\begin{acknowledgements}
We thank M. Gillon for providing information about the TRAPPIST filters, and T. Mikal-Evans and D.K. Sing for useful exchanges about WASP-121b emission spectrum. V.B. and R.A acknowledge support by the Swiss National Science Foundation (SNSF) in the frame of the National Centre for Competence in Research ``PlanetS''. This project has received funding from the European Research Council (ERC) under the European Union’s Horizon 2020 research and innovation programme (project Four Aces, grant agreement No 724427; project Exo-Atmos, grant agreement no. 679633). This paper includes data collected by the TESS mission, which are publicly available from the Mikulski Archive for Space Telescopes (MAST). Funding for the TESS mission is provided by NASA's Science Mission directorate. N.A-D. acknowledges the support of FONDECYT project 3180063. Resources supporting this work were provided by the NASA High-End Computing (HEC) Program through the NASA Advanced Supercomputing (NAS) Division at Ames Research Center for the production of the SPOC data products. This research made use of  {\tt Astropy} \citep{Astropy2013}, {\tt Matplotlib} \citep{Hunter2007} and {\tt Numpy} \citep{vanderWalt2011}, {\tt Scipy} \citep{Jones2001} and {\tt lightkurve} \citep{Barentsen2018}.
\end{acknowledgements}

\bibliographystyle{aa} 
\bibliography{biblio} 

\begin{appendix}

\section{Fit to TESS photometry}
\label{apn:priors_TESS}

\begin{table*}
    \centering
    \caption{Priors for the fit to TESS WASP-121 photometry, based on results from previous analysis by \citealt{Delrez2016}. The notation $\mathcal{N}(\mu,\sigma^2)$ correspond to a normal distribution of mean $\mu$ and variance $\sigma^2$, while $\mathcal{U}(a,b)$ corresponds to a uniform distribution of lower bound $a$ and higher bound $b$.}  
    \label{tab:TESS:priors}
    \begin{tabular}{l|c|c|c} 
        \hline
        \hline
        Parameter & Symbol & Prior & Units \\
                \hline
                \hline
        Orbital period & $P$ &$\mathcal{N}(1.2749248, 7\cdot 10^{-7})$ & d \\
        Transit epoch (BJD$_\text{TDB}$ - 2457000) & $T_{0}$ &$\mathcal{N}(1119.7207, 0.0003)$ & BJD$_\text{TDB}$ \\
        Planet-to-star radii ratio & $R_\mathrm{p}/R_{\star}$ &$\mathcal{N}(0.1234, 0.0050)$ &  \\
        Scaled semi-major axis & $a_\mathrm{p}/R_{\star}$ &$\mathcal{N}(3.82, 0.01)$ &  \\
        Orbital inclination & $i_\mathrm{p}$ &$\mathcal{N}(88.9, 1.7)$ & deg \\
        Argument of periastron & $\omega$ &$\mathcal{N}(9.8, 10.0)$ & deg \\
        Eccentricity & $e$ & $\mathcal{U}(0., 1.)$ &  \\
        \hline 
        Nightside temperature & $T_\mathrm{N}$ &$\mathcal{U}(0., 5000.)$ & K \\
        Sub/antistellar points temperature contrast & $\Delta T_{\rm DN}$ &   $\mathcal{U}(0., 2000.)$ & K \\
        Radiative to advective timescales ratio & $\xi$ &  $\mathcal{U}(-10., 10.)$ &   \\
        \hline 
        Quadratic limb-darkening coefficients & $u_1$ &$ \mathcal{U}(0,1)$ &  \\
        & $u_2$ & $\mathcal{U}(0,1)$ &  \\
        \hline
        Flux scaling coefficient & $C_{\rm s}$ & $\mathcal{U}(0., 1.)$ &  \\
        \hline
    \end{tabular}
\end{table*}

\begin{center}
\begin{figure*}[b!]
\centering
\includegraphics[trim=0cm 0cm 0cm 0cm, clip=True, scale=0.25]{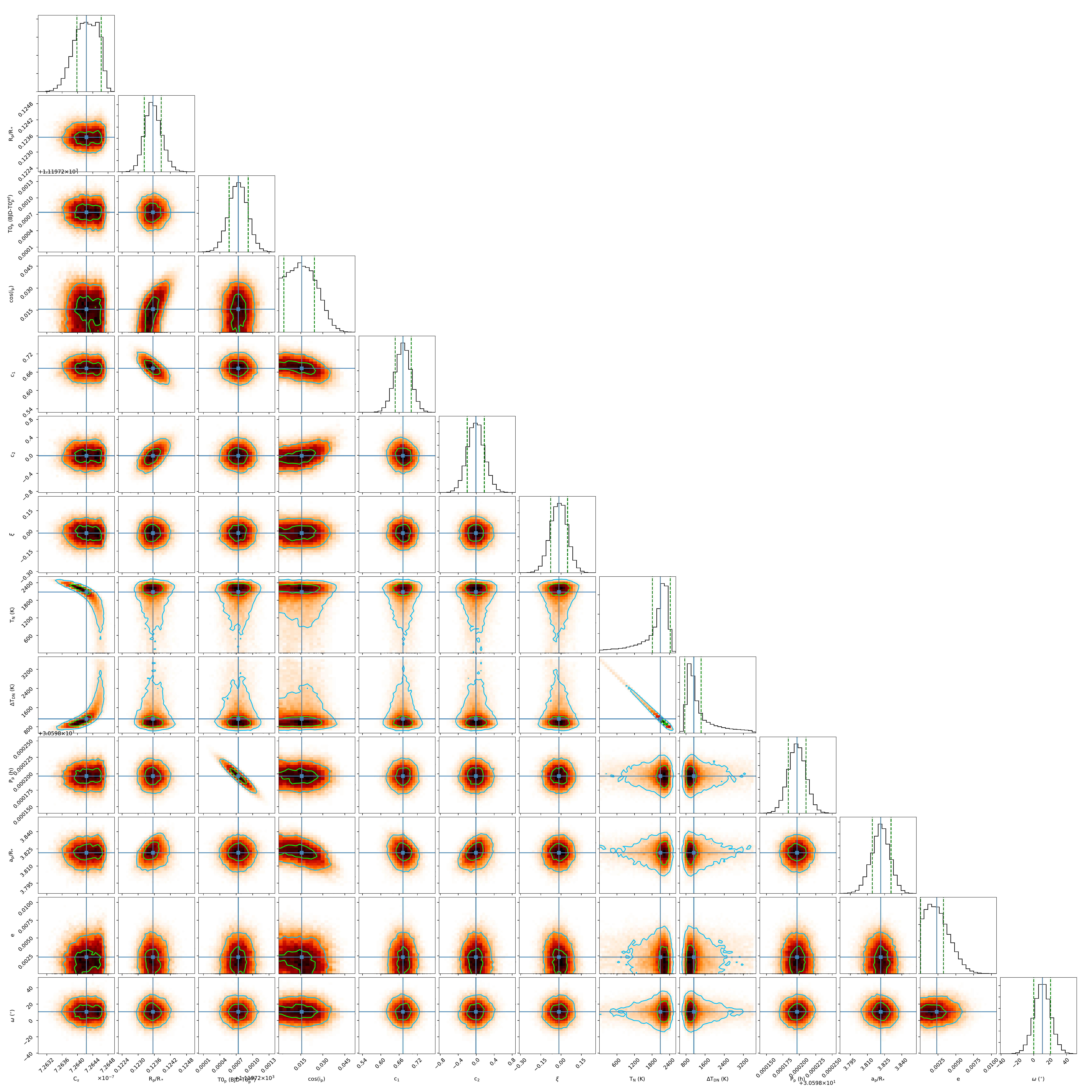}
\caption[]{Correlation diagrams for the probability distributions of the jump parameters to the model fitted to the TESS photometry. Green and light blue contours show the 1 and 2$\sigma$ simultaneous 2D confidence regions that contain respectively 39.3\% and 86.5\% of the samples. 1D histograms correspond to the distributions projected on the space of each line parameter. The deep blue lines indicate their median values, with dashed green lines showing the 1$\sigma$ highest density intervals. }
\label{fig:WASP121_triangle}
\end{figure*}
\end{center}

\section{Atmospheric retrieval}
\label{apn:atmo_retr}

\begin{figure*}
	\centering
	\includegraphics[width=\textwidth]{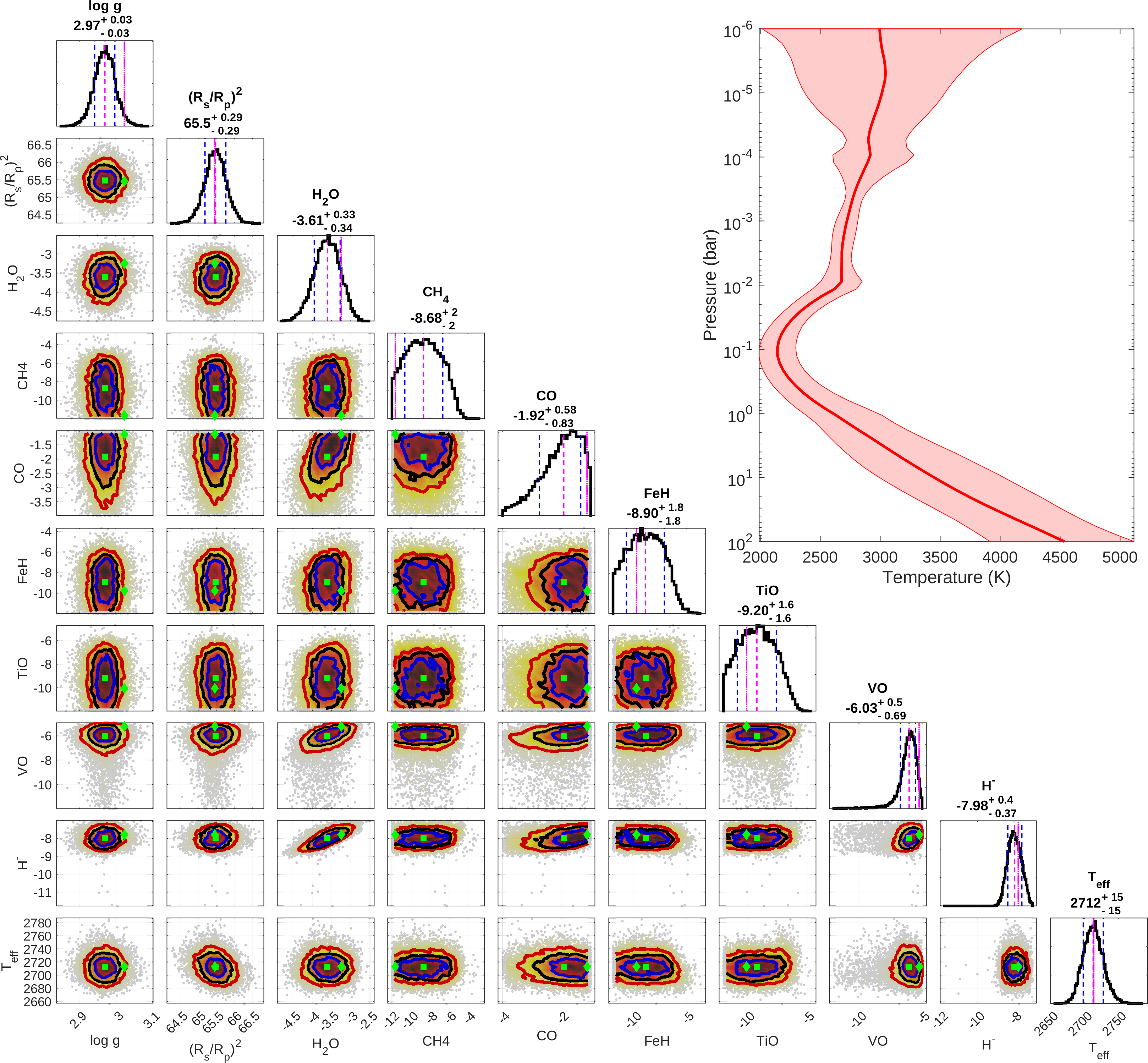}
	\caption[]{Posterior distributions for the retrieval of the WASP-121b emission spectrum using \texttt{Helios-r.2}. The dashed, magenta-colored lines in the posterior plots refer to location of the median value (also stated below each parameter), while the 1\,$\sigma$ confidence limit is denoted by the blue, dashed lines. The magenta, dotted line shows the location of the best-fit model, i.e. the one with highest likelihood value. The molecular abundances are stated in logarithmic units. The solid blue, red, and yellow lines in the two-dimensional parameter correlation plots mark the 1, 2, and 3\,$\sigma$ intervals, respectively. Here, the location of the median (best-fit) model is marked by green squares (diamonds). It should be noted, that T$_\mathrm{eff}$ is not a directly retrieved parameter but a derived quantity. The panel in the upper, right corner depicts the retrieved temperature profile. The solid, red line corresponds to the median profile, while the shaded, red area corresponds to the 1\,$\sigma$ confidence interval.}
	\label{fig:retrival_posterior}
\end{figure*}

\end{appendix}

\end{document}